\documentclass[pra,twocolumn,showpacs,preprintnumbers,amsmath,amssymb]{revtex4}
\usepackage{graphicx}% Include figure files
\usepackage{dcolumn}% Align table columns on decimal point
\usepackage{bm}% bold math
\usepackage{trfsigns}
\newcommand \be{\begin{eqnarray}}
\newcommand \ee{\end{eqnarray}}
\newcommand \ba{\begin{align}}

\newcommand \s{{\backslash\hspace{-1ex}i%\O
}}
\newcommand \2{{\backslash\hspace{-1ex}2%\O
}}

\begin{document}
%\twocolumn[\hsize\textwidth\columnwidth\hsize
%           \csname @twocolumnfalse\endcsname
\title{Bose condensation of squeezed light}
\author{K. Morawetz$^{1,2}$
}
\affiliation{$^1$M\"unster University of Applied Sciences,
Stegerwaldstrasse 39, 48565 Steinfurt, Germany}
\affiliation{$^2$International Institute of Physics- UFRN,
Campus Universit\'ario Lagoa nova,
59078-970 Natal, Brazil}
%\affiliation{$^{3}$ Max-Planck-Institute for the Physics of Complex Systems, 01187 Dresden, Germany
%}
%\affiliation{$^4$Faculty of Mathematics and Physics,
%Charles University, Ke Karlovu 3, 12116 Prague 2, Czech Republic}
%\affiliation{$^5$Institute of Physics, Academy of Sciences,
%Cukrovarnick\'a 10, 16253 Prague 6, Czech Republic}
\begin{abstract} 
Light with a chemical potential and no mass is shown to possess a general phase-transition curve to Bose-Einstein condensation. This limiting density and temperature range is found by the diverging in-medium potential range of effective interaction. The inverse expansion series of the effective interaction from Bethe-Salpeter equation is employed exceeding the ladder approximation. While usually the absorption and emission with Dye molecules is considered, here it is proposed that squeezing can create also such a mean interaction leading to a chemical potential. The equivalence of squeezed light with a complex Bogoliubov transformation of interacting Bose system with finite lifetime is established with the help of which an effective gap is deduced where the squeezing parameter is related to an equivalent gap by $|\Delta(\omega)|={\hbar \omega/( \coth {2|z(\omega)|}-1)}$. This gap phase creates a finite condensate in agreement with the general limiting density and temperature range. In this sense it is shown that squeezing induces the same effect on light as an interaction leading to possible condensation. The phase diagram for condensation is presented due to squeezing and the appearance of two gaps is discussed. 
\end{abstract}

\pacs{
}
\maketitle

\section{Introduction}

Despite being a figure of jokes in short stories, the storing of light in boxes has become reality. Infinite fluctuations in numbers of photons prevent such capturing of light. This impossibility has been recently circumvented by sufficient high absorption and emission rates which provides thermal equilibrium and renders the chemical potential finite. This is achieved by photonic bandgapped materials whose Bragg structures were predicted to trap, store, and release light \cite{YKBS05,YKBS05a} or to show pairing in nonlinear polar crystals \cite{Ch91}. The Bose-Einstein condensation of light now can be used to down-convert the frequency of light \cite{Na03}.   

Polaritons as hybrid states between excitons and polarons show trapped light condensation \cite{RKRAD05,Kasprzak2006b,BHSW07,KSADSM08} and
the condensation of excitons is measured in potential traps \cite{butov02}. Possible transitions to an electron-hole liquid formation are discussed in \cite{keldysh97}. The Bose-Einstein condensation of excitons in bilayer quantum Hall structures \cite{Ass11} is only stable at a small ratio of a quantum well distance to the magnetic length \cite{doretto06}. The effect of quantum confinement on excitons  is investigated in quantum dots of indirect-gap materials \cite{takagahara92}. In single heterostructures the size quantization is absent and e-h interaction with their images can cause exciton localization near the contacting medium surface creating of presurface excitons \cite{tkach96}. Giant permanent dipole moments of excitons are found in semiconductor  nanostructures \cite{warburton02} and dipolaritons with electric dipole moment in tunnel-coupled
quantum wells have been investigated in \cite{SKYYD14}. 

These results of equilibrium phase transitions should be taken with care, however. There are more effects one might think about. In two-dimensional systems the condensation by interactions between excitons as a system of islands in an exciton gas can explain the condensed phase in quantum wells and periodical fragmentation measured in luminescence without the requirement of Bose-Einstein condensation \cite{sugakov06}. Also out-of-equilibrium Bogoliubov modes have been observed \cite{Ass11} and no thermodynamic
equilibrium phase transition has been found though features from a Bose-Einstein condensation are verified. Instead, polariton lasing has been reported \cite{KF10} at room temperature in an organic single-crystal microcavity and thermalization of polaritons and condensation has been seen also in polymers \cite{Plumhof2014}. This shows that the range of lasing and of Bose-Einstein condensation is most probably a nonequilibrium balance.
For a comprehensive overviews of polariton condensates at room temperature and the regime of polariton lasing \cite{SCH13} see \cite{DHY10,Keeling2011,GB16}.

Let us return to the Bose-Einstein condensation of light which has been observed finally in a two-dimensional photon gas confined in a box with dye molecules \cite{Klaers2010a} where the chemical potential was freely adjustable \cite{KVW10,KSDVW11}. The confining potential by the cavity mirror provides an effective photon mass \cite{KSDVW12} such that the analogy with the ideal Bose gas is suggestive. Also further observed properties underline this interpretation. A cusp-like singularity in the specific heat was reported by calorimetric measurements in \cite{DSLDVWK15}. The {\em in situ} observation of the crossover from laser-driven dissipative system to a thermalized gas was seen in \cite{SDDVKW15}. A coherent coupling between a laser-driven optomechanical membrane and a Bose-Einstein condensate has been achieved as well \cite{FN17,DNM18} which shows strong squeezing in the mechanical mode \cite{DNM18}. The optical cavity has been used to demonstrate variable potentials leading to a small critical photon number of $N_0=68$ for condensation \cite{DKDSVWK17} which physical principles are discussed in \cite{NW18}. 

Though Bose-Einstein condensation of light has been seen due to the effective indirect interaction of light with the Dye molecules, superfluidity is not yet observed though the Bogoliubov excitation spectrum has been calculated and the conditions for superfluidity has been investigated \cite{ASBPW17}.
A possible superfluid-Mott-insulator transition of a two-dimensional photon gas in the presence of a periodic potential has been suggested in \cite{LODS15}.
The propagation
of quantum light and its thermalization towards condensation 
has been investigated in 
\cite{CLC16}
to obtain spontaneous macroscopic optical coherence. 

Here the same caution is necessary given that it is a highly nonequilibrium process. Since the pump power for condensation depends
on the pump beam geometry and the cavity cutoff wavelength, it suggests that energy-dependent thermalization and loss mechanisms are important \cite{MN15}. 
Moreover, a decondensation under nonequilibrium conditions has been reported in \cite{HNRM18} which reminds that the seemingly equilibration due to sufficient emission and absorption of light is a delicate balance leading to lasing or condensation \cite{KK13}. Phase diffusion
of a Bose-Einstein condensate of photons has been considered \cite{LWDS14} where a corresponding interference experiment was proposed. The finite-lifetime effects on first-order correlation functions of dissipative Bose-Einstein condensates are important in this respect \cite{LSD14} and the interactions on condensate-number fluctuations are treated in \cite{WLDS14}. A mixed gas of photons and photon pairs has been reported finally in \cite{ZYZ12}. We will see finite lifetime effects in our model.

Naturally these experimental progress has triggered an enormous theoretical activity. We want only briefly mention some lines of ideas.
Spatial and temporal coherence of condensed microcavity polaritons have been
calculated in terms of the Boltzmann equation \cite{DCTTH08}. Disorder
can increase here the light-matter coupling \cite{GP08}. Condensation of Bosons interacting only
with incoherent phonons and spontaneous amplification of quantum coherence are theoretically reviewed in \cite{Snoke2013}. With a two-level model of gaseous medium the effective mass due to light trapping has been employed to examine the influence
of intra-cavity medium on the parameters of light condensation \cite{Kru14} and for a one-dimensional condensate in a microtube \cite{Kru16}. Generalized superstatistics by the maximum entropy principle was applied to fluctuations of the photon Bose-Einstein
condensate in a dye microcavity \cite{Sob13}. 
The linewidth of a single-mode polariton condensate was calculated by a quantum jump approach in the wave-function Monte Carlo method \cite{Wou12}.

All these approaches rely on the capturing of light in a microcavity creating an effective mass and a chemical potential. In principle only an effective interaction of photons with the surrounding, like Dye molecules, is sufficient to produce an effective chemical potential and to show consequently Bose-Einstein condensation. Here we will follow this path and consider light with no effective mass and therefore linear dispersion and will show that a finite chemical potential can be realized by a proper squeezing of light. An equivalence of squeezing and complex Bogoliubov transformation will be presented analogously to the squeezing Bogoliubov automorphisms \cite{HR96,Honegger1998,BD07}. With the complex Bogoliubov transformation we introduce an additional degree of freedom as a phase while in \cite{YY02} the Bogoliubov transformation was generalized by adding a term. With the help of such a transformation we will find that the squeezed light can be considered as a Bose gas of finite lifetime with a gapped spectrum as it would follow from interaction correlations. In this sense we show that squeezing behaves in the same way as making the light interacting. Therefore the paper consists of two parts, first to show that light exhibits possible Bose-Einstein condensation if it is interacting and second, to show that squeezing realizes the necessary interaction correlations.  

Our approach is motivated by the observation that the condensate mode is squeezed \cite{HR06} where the squeezing parameter scales with the system size to have no thermodynamic effect. By variational calculation of the ground state wave function, the pair state has been found to be a multimode squeezed vacuum state \cite{M91}. Nonlinear effects of atomic collisions are found being similar to degenerate parametric amplifies and have been used to manipulate the condensate mode and to produce squeezed states \cite{DNS14}. Here in this paper we will employ squeezing to create such a condensed state. 

The outline of the paper is as follows. In the next chapter we present a general borderline of a possible phase transition of light with an ad-hoc assumed effective chemical potential but no effective mass. It will be found that an interaction created by any coupling to the environment leads unavoidable to a phase transition to Bose-Einstein condensates. We will employ the inverse series expansion of the effective range by the Bethe-Salpeter equation. In this way the ladder approximation of the ${\cal T}$-matrix is exceeded. In chapter III we present the equivalence of squeezed light and a complex Bogoliubov transformation which allows to consider squeezed light as an interacting gas of Bosons with finite lifetime. The theoretical forms of occupation numbers are found to be exactly the ones resulting from the multiple-scattering corrected ${\cal T}$-matrix which is shortly remembered in chapter IV. This allows to identify an effective gap in chapter V where a model of the frequency dependencies of squeezing and gap parameters is introduced. The resulting phase diagram is presented in chapter VI. Finally we summarize and conclude in chapter VII. In the appendix we briefly sketch the multiple-scattering ${\cal T}$-matrix approach which allows to describe interacting systems in both phases, in and out of the condensate, on the same footing.  

\section{Phase diagram of interacting photons}\label{pha}
Let us start with the assumption that there is a mechanism which creates a selfenergy $\Sigma_0=T\sigma>0$, e.g. by a meanfield due to Dye molecules or as we will see in the next chapter by squeezing.
We consider this as interacting light with the linear dispersion $\varepsilon=pc+\Sigma_0$ where the pole part of the selfenergy $\Sigma_0$ can be interpreted as effective negative chemical potential, $\mu^*=-\Sigma_0$. Any mechanism leading to a selfenergy creates such an effective chemical potential which renders the number of photons finite
\be
n=\int{d^3 p\over (2 \pi\hbar)^3}{1\over {\rm e}^{pc-\mu^*\over T}-1}={T^3\over \pi^2\hbar^3 c^3}P_3\left ({\rm e}^{-\sigma}\right ).
\label{n}
\ee
Here and in the following we use the polylog function $P_n(z)=\sum_{k=1}^\infty{z^k/ k^n}$. Like in the normal non-interacting Bose gas the critical temperature is reached when the Bose pole is approached for $\mu^*\sim -\sigma\to 0$, i.e.
\be
T_{c1}=\pi^{2/3}\xi_3^{-1/3} \hbar c  n^{1/3}
\label{T0}
\ee
where $\xi_3=P_3(1)$.
This critical temperature changes due to higher-order approximations of the
interaction correlations \cite{BBHLV99,Hu99,AMT01,CPR01,BBHLV99,MMS07,BSP15}. 
Bose-Einstein condensation for a general dispersion $\varepsilon=c_s p^s$ are discussed in \cite{ALS99} where it has been found that the critical temperature exists only for dimensions larger than $s$. We consider linear dispersion such that a critical temperature in three dimensions is possible.

A general note might be in order. Bose condensation appears when the chemical potential reaches the bottom of dispersion and any extra particle has to enter the condensate. Here we can assume that the bottom of linear dispersion is at zero. Therefore a zero chemical potential is not equivalent to an undetermined photon number itself. A fixed photon number in and out of condensate is present if we have a mechanism which creates a chemical potential at all, such that the relation between density and chemical potential (\ref{n}) exists. This will be shown by interactions and in the second part by squeezing. Let us start with interacting light and the Hamiltonian 
\be
H=\sum\limits_p \epsilon_p a_p^+a_p+\frac 1 2 \sum\limits_{kpq}V_{{k-p\over 2},{k-p\over 2}+q} a_{k+q}^+a_{p-q}^+a_pa_k
\ee
with Bosonic creation $\hat a^+$ and annihilation $\hat a$ operators, linear dispersion $\epsilon_p=c p$, and assuming a separable interaction $V_{pp'}=\lambda g_pg_{p'}$ with e.g.
a Yamaguchi \cite{Y59} form factor $g_p=1/(p^2+\beta^2)$. For other form factors see \cite{MSSFL04}.
We will derive the medium-dependent scattering range from solving the Bethe-Salpeter equation which diverges near the phase transition. This method has been used before in \cite{MMS07,SMR97} such that we might present it here only briefly.

The Bethe-Salpeter equation for
the many-body ${\cal T}$-matrix with a center-of-the-mass momentum $K$, and incoming and
outgoing difference momenta $p$ and $p'$ of the two particles, respectively, can be solved 
\ba 
{\cal T}_{pp'}(K,\omega,\mu)&\!=\!V_{pp'}\!+\!\int \!\!{d^3 q\over (2 \pi \hbar)^3}V_{pq}{1\!+\!f_{\frac K 2 \!+\!q}\!+\!f_{\frac K 2 \!-\!q}\over \omega\!-\!\epsilon_{\frac K 2 \!+\!q}\!-\!\epsilon_{\frac K 2 \!-\!q}\!+\!i0}T_{q,p'}
\nonumber\\
&={\lambda g_p g_{p'}\over 1-\lambda J(K,\omega,\mu)}
\label{T}
\end{align}
with the Bose distribution $f_p$ and
\be
J(K,\omega,\mu)=\int {d^3 q\over (2 \pi \hbar)^3}g_{q}^2{1+f_{\frac K 2 +q}+f_{\frac K 2 -q}\over \omega-\epsilon_{\frac K 2 +q}-\epsilon_{\frac K 2 -q}+i0}.
\label{J}
\ee
The in-medium scattering phase shift is given by the on-shell ${\cal T}$-matrix at $K=0$
\ba
\tan{\phi}&=\left . {{\rm Im} {\cal T}\over {\rm Re} {\cal T}}\right |_{\omega=c p}
%\nonumber\\
={-\pi p^2 g_p^2 (1+2 f_p) \lambda\over 4 \pi^2 \hbar^3c-\lambda \int\limits_0^\infty d q {q^2g_q^2\over p-q} (1+2 f_q)}
\nonumber\\
&= {a p\over \hbar}+{r^2 p^2\over 2 \hbar^2}+o(p^3)
\end{align}              
which expansion shows that the scattering length $a=0$ vanishes for the linear dispersion $\varepsilon_p=cp$ here and the in-medium scattering range $r$ reads
\be
r^2={-2\pi \hbar^2 g_0^2(1+2 f_0)\over {4\pi^2\hbar^3 c\over \lambda}+\int\limits_0^\infty dq q g_q^2(1+2 f_q)}.
\label{r}
\ee 
We can observe \cite{SMR97} that the interaction is behaved such that the denominator vanishes itself when approaching the phase transition which means
\be
{1+2 f_0\over r^2}\sim 0\quad {\rm for}\,\, \sigma\to 0.
\label{cond}
\ee
The medium-free expression $r_0$ appears if we set $f_p=0$ in (\ref{r}) which we consider as the experimental given value. Therefore we express the coupling constant $\lambda$ in terms of this variable $\lambda(r_0)$ with the help of (\ref{r}) and accordingly eliminate $\lambda$ in (\ref{T}). In such a way the ${\cal T}$-matrix becomes renormalized without explicit cut-off which corresponds to a running coupling constant. Near the phase transition the in-medium scattering range $r^2$ will diverge and we can use the ratio to the medium-free scattering range as a measure of such phase transition 
\be
\left ({r\over r_0}\right )^2={1+2 f_0\over 1-{r_0^2\over \pi \hbar^2}\int\limits_0^\infty q {g_q^2\over g_0^2} f_q}.
\label{rr0}
\ee
In the following it is sufficient to consider the limit of contact potential $\beta\to\infty$ which renders $g_q/g_0\to 1$. Please note that due to the running coupling constant we are introducing a modified contact potential allowing also a repulsive behaviour which is impossible for pure contact potentials \cite{M09}. 

Using the density $n$ we can define the dimensionless in-medium order parameter $y=n^{1/3} r$. With the help of (\ref{rr0}) we then express the medium-free $y_0=n^{1/3} r_0$ in terms of the medium one as
\be
y_0^2={y^2\over \coth{\sigma\over 2}-\pi^{1/3} g'(\sigma) y^2}\to -{1\over \pi^{1/3} g'(\sigma)}
\label{y0}
\ee 
for $y\to\infty$ when approaching $\mu^*=-T\sigma\to 0$ where the Bose function diverges.
While $y_0$ is free of any Bose function $f_0$, the medium-dependent $y= n^{1/3} r$ of (\ref{r}) diverges at this limit. Since the term $\coth \sigma/2=1+2 f_0$ diverges slower than $y^2$ according to (\ref{cond}) we see that the limit in ({\ref{y0}) corresponds indeed to $\mu^*\to 0$ and therefore describes the curve of phase transition. Furthermore $g'(\sigma)$ is the derivative of 
$g(\sigma)=3 P_3^{1/3}({\rm e}^{-\sigma})$ with $\sigma=-\mu^*/T$ and $P_3'({\rm e}^{-\sigma})=-P_2({\rm e}^{-\sigma})$ from (\ref{n}). 
The $\sigma$ is the coupling  of the order parameter to an external bath due to the selfenergy $\sigma=\Sigma_0/T$. 

Next, we can express the ratio of the temperature to the medium-free critical one (\ref{T0}) according to (\ref{n})
\be
x={T_c\over T_{c1}}={g(0)\over g(\sigma)}
\label{x}
\ee  
and obtain how the critical temperature of phase transition changes with the coupling $\sigma$. In principle, from (\ref{y0}) and (\ref{x}) we can now create a simple parametric plot $\{x(\sigma),y_0(\sigma)\}$ to see how the critical temperature is changing in terms of the order parameter $y_0=n r_0^3$ which is the density times scattering range $r_0$. Systems with spontaneous symmetry breaking show bifurcations in order parameters even for vanishing external perturbation. The simple elimination of $\sigma$ in $x$ and in $y_0$ does not provide such symmetry breaking since we expand directly with respect to the perturbation $\sigma$. This situation is changed completely if we work with the inverted series $y_0(\sigma)\rightarrow \sigma(y_0)$ since then a vanishing perturbation $\sigma$ does allow multiple solutions of the order parameter $y_0(x)$ which we search for as signal of phase transition. This method is known as inversion method and described thoroughly in \cite{FKYSOI95}.

\begin{figure}[h]
  \includegraphics[width=8cm]{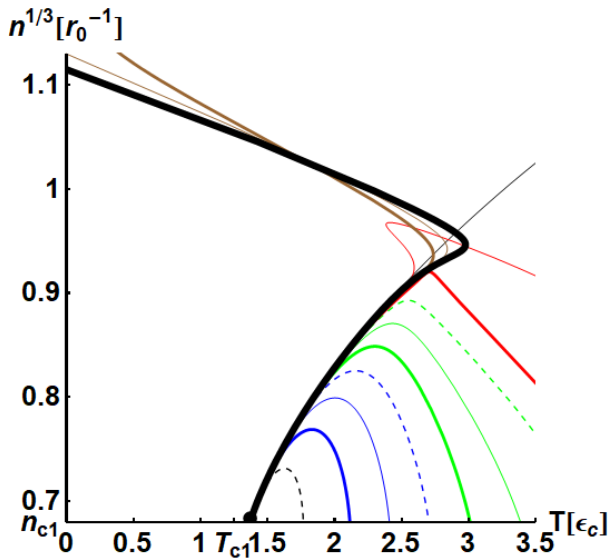}
\caption{The critical density times interaction range versus critical temperature for 2d-14th order in the expansion $\sigma(y_0)$ (right side from below to above). The simple parametric plot $\{x(\sigma),y_0(\sigma)\}$ is shown as continuously increasing thin black line to the upper right. Below $T_{c1}$ and $n_{c1}$ the linear dependence (\ref{T0} holds. The scaling is (\ref{ec}).
}
\label{Tc}
\end{figure}

In figure \ref{Tc} we plot this expansion together with the simple parametric plot $\{x(\sigma),y_0(\sigma)\}$. While this plot shows a continuous increase of $y_0$ with $x=T_c/T_{c1}$ the inverse expansion approximates this curve up to the 9th order from the right. Starting with the 10th-order expansion the curve begins to bend to the left converging visibly around the 14th order. Therefore we see the appearance of exactly the above described symmetry breaking by a back-bending of the curve and the observation that one critical temperature is realized by two different densities. Remarkably the critical temperature increases first with increasing density $y_0=n^{1/3} r_0$ according to (\ref{T0}) and decreases above the maximum of $x=T_c/T_{c1}=1.45$ at 
\be
n_{\rm max} r_0^{1/3}\approx 0.9445
\ee
to vanish at the upper critical density of photons of 
\be
n_{\rm c2}\, r_0^{1/3}\approx 1.1539
.
\ee
Below the temperatures $T_{c1}$  given by (\ref{T0}) and below the density 
\be
n_{\rm c1} r_0^{1/3}={\sqrt{6} \xi_3^{1/3}\over \pi^{7/6}}=0.685029
\ee
the usual linear dependence (\ref{Tc}) is present.  As typical energy and temperature scale it is used further-on 
\be
\epsilon_c={\hbar c\over r_0}=2.300 {\rm K}\, k_B{{\rm mm}\over r_0}=0.1973\, {\rm eV}\,{{\rm \mu m}\over r_0}
\label{ec}
\ee 
with the Boltzmann constant $k_B$. This means an effective range $r_0$ in orders of 10th of micrometer would lead to room temperatures. The second expression in (\ref{ec}) provides the energy scales we are dealing with.

%Remembering the critical temperature of the medium-free gas (\ref{T0}) we can plot also the absolute critical temperature in figure \ref{Tcc}.
%\begin{figure}[]
%  \includegraphics[width=8cm]{Tcc1.eps}
%\caption{The interaction parameter versus absolute critical temperature for the curve of figure~\ref{T0}. The linear dependence (\ref{Tc}) is plotted as dashed line and the parametric plot as thin line. 
%}
%\label{Tcc}
%\end{figure}

\section{Squeezed light and complex Bogoliubov transformation}
Now we will investigate a possibility to create Bose condensation of light without the recourse to Bose systems with finite chemical potential. This will turn out to be squeezed light which we will demonstrate to create effects like originating by interactions. In this sense we will realize the necessary interaction correlations of the foregoing chapter by squeezing. Squeezing is realized by the two-particle squeezing operator 
\be
\hat S={\rm e}^{{z\over 2} (\hat a^+)^2-{\bar z\over 2} (\hat a)^2}
\ee
with the complex squeezing parameter
\be
z=|z|{\rm e}^{i\phi}.
\label{z}
\ee 
The coherent light can be considered as produced by an analogous one-particle displacement operator which leads just to a shift in the densities \cite{BaR02} and which therefore we do not need here explicitly.

Employing the commutator relation
\be
\hat S \hat a^+=\left [a^+\cosh(|z|)-\hat a {z\over |z|} \sinh(|z|)\right ]S^+,
\ee
the mean density of squeezed light reads \cite{BaR02}
\ba
n_g(p)&={\rm Tr} \hat S^+\rho \hat S \hat a^+\hat a
={\rm Tr}\rho [ \hat a^+\hat a \cosh^2{|z|}+\hat a\hat a^2 \sinh^2{|z|}]
\nonumber\\
&=f_p\cosh^2(|z|)+(1+f_p) \sinh^2 (|z|)             
\label{sn1}
\end{align}
with the Bose distribution $f_p={\rm Tr} \hat a^+_p\hat a_p$.
Analogously we find the anomalous density
\be
n_-\!=\!{\rm Tr} \hat S^+\rho \hat S \hat a\hat a=\!-\!{z\over |z|} \sinh(|z|)\cosh(|z|) (1\!+\!2 f_p).
\label{sn2}
\ee
Now we observe that the occupation numbers (\ref{sn1}) and (\ref{sn2}) of this squeezed light have exactly the form which appears when we perform a complex Bogoliubov transformation 
\be
\hat a&=&u\hat \alpha-v \hat \beta^+
\nonumber\\
\hat b&=&u\hat \beta-v \hat \alpha^+
\label{bog}
\ee
of the Hamiltonian
\ba
\hat H&=\epsilon_0(\hat a^+\hat a\!+\!\hat b^+\hat b)+\epsilon_1 (\hat a^+\hat b^+\!+\!\hat a \hat b)
\nonumber\\
&=[\epsilon_0(|u|^2\!+\!|v|^2)-2 \epsilon_1 \cos[\phi_u\!+\!\phi_v)|u||v|](\hat \alpha^+\hat \alpha\!+\!\hat \beta^+\hat \beta)
\label{H}
\end{align}
with $\hat b_k=\hat a_{-k}$. Then one introduces the new Bosonic operators $\hat \alpha$ and $\hat \beta$ which 
have the mean value of the quasiparticle distribution
\be
f_b={\rm Tr}\hat \rho \hat \alpha^+\hat \alpha={\rm Tr}\hat \rho \hat \beta^+\hat \beta={\rm Tr}\hat \rho \hat \alpha^+\hat \beta={\rm Tr}\hat \rho \hat \beta^+\hat \alpha
\ee
and one obtains as mean density distributions
\be
n_g(p)&=&u^2 f_p+v^2(1+f_p)\nonumber\\
n_-(p)&=&-u v (1+2 f_p).
\label{np}
\ee
Comparing with (\ref{sn1}) and (\ref{sn2}) allows to identify
the modulus and phase
\be
|u|=\cosh(|z|),\quad|v|=\sinh(|z|),\quad \phi=\phi_u+\phi_v.
\label{uv}
\ee
The complex $u$ and $v$ can model a complex dispersion of the energy $\epsilon_0=\epsilon_{0r}+i\epsilon_{0i}$ as well as a complex interaction $\epsilon_1=\epsilon_{1r}+i\epsilon_{1i}$. Examples are the dispersion in the presence of thermo-optic photon interactions, see figure 4 of \cite{ASBPW17}, or the damping due to saturation of molecular excited states \cite{NW18}. Also the lifetime effects considered in \cite{LSD14} can be thought to be parameterized this way.

In other words the squeezing is shown to lead to the same occupations as a Bogoliubov transformation of the complex-valued Hamiltonian (\ref{H}) concerning exclusively the occupation. This does not mean that any realization of squeezed light is actually evolving under the Hamiltonian (\ref{H}). Only concerning the occupation (\ref{sn1}) and (\ref{sn2}) of any squeezed light we have shown that the effective Hamiltonian (\ref{H}) is equivalent to such realization. Some specific experimental recipes might be implementing this Hamiltonian directly.

In terms of the squeezing parameter (\ref{z}) we find the identities
\be
\epsilon_{0i}&=&-\frac 1 2 \sin{(2\phi)}{\sinh^2{(2|z|)}\over\cosh{(2|z|)}}\epsilon_{0r}
\nonumber\\
\epsilon_{1r}&=&\cos{(\phi)}{\tanh^2{(2|z|)}}\epsilon_{0r}
\nonumber\\
\epsilon_{1i}&=&-\sin{(\phi)}{\sinh^2{(2|z|)}}\epsilon_{0r}
\label{e0i}
\ee
uniquely determined by the demand that the quasiparticle dispersion (\ref{H}) should be real. Indeed, we obtain after the Bogoliubov transformation (\ref{bog}) the Hamiltonian (\ref{H}) in diagonal form 
\be
\hat H=E (\hat \alpha^+\hat \alpha+\hat \beta^+\hat \beta)
\ee
with alternatively
\be
E&=&\sqrt{(\epsilon_{0r}+i\epsilon_{0i})^2-(\epsilon_{1r}+i\epsilon_{1i})^2}
\nonumber\\
&=&\epsilon_0(|u|^2+|v|^2)-2 \epsilon_1 \cos[\phi_u+\phi_v)|u||v|
\nonumber\\
&=&\epsilon_{0r}\left [\cosh{(2|z|)}-\cos^2{(\phi)}{\sinh^2{(2|z|)}\over \cosh{(2|z|})}\right ]
\nonumber\\
&=&\epsilon_{0r}\left [\sin^2{(\phi)}\cosh{(2|z|)}+{\cos^2{(\phi)}\over \cosh{(2|z|)}}\right ] >0
\label{E}
\ee
which implies $\epsilon_{1i}\epsilon_{1r}=\epsilon_{0i}\epsilon_{0r}$. 
We have shown therefore that the squeezing induces densities which result usually from interacting Hamiltonians of the form (\ref{H}). We might consider the Hamiltonian (\ref{H}) as a representation of some specific experimental recipe to produce squeezed light in cavities, e.g. by continuously driving a nonlinear crystal in a lossy cavity.

Please note that similar Bogoliubov operators have been used to solve a two-photon quantum Rabi model exactly leading to extended squeezed states \cite{CWHLW12}. The Bogoliubov theory was also used to create a spin-squeezed entangled state \cite{Soe02}. The conditions on the parameters to guarantee unitarity for relativistic charged particles are worked out in \cite{Ru77}. 

In order to establish the link to the standard treatment of gapped phases within the theory of interacting Bose gases we rewrite (\ref{uv}) to identify the 'gap' form
\be
{\Delta (\omega)\over \sqrt{(\hbar \omega+|\Delta (\omega)|)^2-|\Delta (\omega)|^2}}&=&{\rm e}^{i\phi} \sinh{2 z(\omega)}
\nonumber\\
{\hbar \omega+|\Delta (\omega)|\over \sqrt{(\hbar \omega+|\Delta (\omega)|)^2-|\Delta (\omega)|^2}}&=& \cosh{2 z(\omega)}
\ee
which means we have introduced the gap
\be
|\Delta(\omega)|={\hbar \omega\over \coth {2|z(\omega)|}-1}.
\label{gap}
\ee
This is an exact rewriting but establishes a new link of a squeezing parameter to the equivalent description by a gap phase.  
The density distribution (\ref{np}) becomes with the help of (\ref{uv})
\ba
n_g(\omega)={1\over2}\left [{\epsilon(\omega)\over E(\omega)}\left (1\!+\!2 f_{E(\omega)}\right )-1\right ]
\label{fg}
\end{align}
with the quasiparticle energy
\be
E(\omega)= \sqrt{\varepsilon(\omega)^2-|\Delta (\omega)|^2}
\label{E1}
\ee
and the free energy
\be
\varepsilon(\omega)=\hbar \omega-\mu^*.
\label{eps}
\ee
Here we denoted the effective chemical potential $\mu^*=\mu-\Sigma_0$ with some selfenergy $\Sigma_0$ at zero frequency. The original chemical potential $\mu$ might be zero as in free light. The critical line is given by $\mu^*=-\Delta$ and $\Delta_0\ne 0$ due to squeezing.

A condensate appears if a macroscopic number of photons is in the ground state
$N_0=n_g(0)$. Eq. (\ref{fg}) for $T\ll E(0)$ leads then to an equation for the chemical potential
\be
\mu^2-2 \Sigma_0\mu+\Sigma_0^2-\Delta_0^2=o(N_0^{-1}).
\ee
For macroscopic number of photons in the condensate $N_0\to \infty$, this provides just the Hugenholtz-Pines theorem \cite{HP59}
\be
\mu^*=\mu-\Sigma_0=-\Delta_0.
\label{HP}
\ee

\section{Multiple-scattering corrected ${\cal T}$-matrix approach}\label{tcorr}

We have established that squeezing leads to the same correlations as it would follow from a Bogoliubov transformation of a complex-valued Hamiltonian leading to a gap structure. Therefore we can now use the theoretical description of correlated Bose gases exhibiting gap and condensed phases. This is accomplished by the summation of ladder diagrams leading to the many-body ${\cal T}$-matrix.

Bose-Einstein condensation and the density distribution (\ref{fg}) for the appearance of a gap do not follow from the symmetric ${\cal T}$-matrix (\ref{T}). It has been recognised quite early that in order to obtain the gap equation for pairing one needs one selfconsistent and one non-selfconsistent propagator for the intermediate state in the Bethe-Salpeter equation of the ${\cal T}$-matrix \cite{Prange60,KM61}. Later it was used as an ad-hoc approximation seemingly violating the symmetry of equations and consequently violating conservation laws \cite{MBL99,HCCL07}. This has remained puzzling  until it became clear that nonphysical repeated collisions with the same state in the symmetric ${\cal T}$-matrix are the reason. When these repeated collisions are removed from the ${\cal T}$-matrix, the correct gap equation appears and the condensate can be described without asymmetrical ad-hoc assumptions about selfconsistency \cite{L08}. It is a consequence of the hierarchical dependencies of correlations \cite{Mo10}. This scheme of multiple-scattering corrected ${\cal T}$-matrix has been successfully applied to describe the critical behaviour in and outside a condensate in superconductivity \cite{SLMMM11} and in Bose-Einstein condensation \cite{M11_1,MML13}.

The wrong multiple scattering events are visible only in a singular channel, i.e. the condensate, where the occupation becomes macroscopic, since otherwise the weight of a single channel is vanishing in the thermodynamic limit. We subtract this singular channel from the  propagator in order to avoid multiple scattering  
\be
G_\s=G-G_\s\Sigma_iG.
\label{sub}
\ee
Now we consider a general ${\cal T}$-matrix which represents the regular selfenergy as
$\Sigma=\Sigma_{\rm tot}-\Sigma_i=\sum_{j\ne i} {\cal T}_j\bar G$ where the channel ${\cal T}$-matrix ${\cal T}_j$ as
two-particle function is closed by an backward propagator $\bar G$.
In the singular channel we subtract the repeated interaction within this channel and close with the subtracted propagator $\Sigma_i={\cal T}_i\bar G_\s$. The Dyson equation takes then the form \cite{M10}
\be 
G^{-1}&=&G_0^{-1}-\Sigma_{\rm tot}
%\nonumber\\&=&
=G_0^{-1}-\Sigma-\Sigma_i
=G_0^{-1}-\Sigma-{\cal T}_i\bar G_\s
\nonumber\\
&=&G_0^{-1}-\Sigma-{\cal T}_i\left (\bar G_0^{-1}-\bar \Sigma\right )^{-1}.
\label{gm}
\ee
Since the free propagator is $G_0^{-1}=\omega-\epsilon_k$ or $\bar G_0^{-1}=-\omega-\epsilon_{-k}$ and the ${\cal T}$-matrix in the singular channel is $T_i(k)=\Delta_k \bar \Delta_{-k}$, the retarded propagator (\ref{gm}) reads
\ba
G&={\hbar \omega+\epsilon_{-k}+\Sigma_{-k} \over (\hbar \omega+\epsilon_{-k}+\bar \Sigma_{k})(\hbar \omega-\epsilon_k-\Sigma_{k})-\Delta^2_k}
\nonumber\\
&=\frac 1 2 \left (1\!+\!{\varepsilon\over E}\right ) {1\over
  \hbar \omega\!-\!\mu^*\!-\!E}\!+\!\frac 1 2 \left (1\!-\!{\varepsilon\over E}\right ) {1\over
  \hbar \omega\!-\!\mu^*\!+\!E}
\label{polebose}
\end{align}
which gives just the distribution (\ref{fg}) with (\ref{E1}) and (\ref{eps}).
This scheme can be written for Fermi or Bose systems, for details see \cite{M17b}.

The condensation appears at poles of the multiple-scattering-corrected ${\cal T}$-matrix at $\mu^*=-\Delta$ and $\hbar \omega=2\Delta$ according to the Thouless criterion \cite{Thouless60}. We use the separable potential of chapter~\ref{pha} and the running coupling constant renormalization $\lambda(r_0)$ as discussed after (\ref{r}). This inverse ${\cal T}$-matrix reads then for $g_p\to 1$
\be
{2\pi \epsilon_c r_0^3\over {\cal T}(-\Delta,2\Delta)}=\frac {1}{2\pi}\int\limits_0^\infty d\Omega \Omega^2\left [{\coth{E\over 2 T}\over E}-{1\over \Omega}\right ]-1
\label{inverse}
\ee
which is convergent.

An important observation is that for the normal phase (\ref{J}) with $\omega\ne 0$ the renormalized ${\cal T}$-matrix vanishes for formfactors $g_p\to 1$, which means contact interaction, since the frequency integral in the denominator diverges. Only for $\mu=-\Delta=-\omega/2 \to 0$ a finite value results which is the same limit as appearing from (\ref{inverse}). This divergence of the denominator in the normal phase is typical for linear dispersion $\hbar\omega =c p$ since the density of states is $\sim\omega^2$. For quadratic dispersion $\hbar\omega \sim p^2$ the density of states is $\sim \omega^{3/2}$ and the integration would become finite leading to the discussion of bound and scattering states as in the original paper of Yamaguchi form factors \cite{Y59} or other formfactors \cite{MSSFL04}. We observe here that for interacting light the normal ${\cal T}\to 0$ vanishes due to the diverging denominator which means the effective interaction is vanishing in the normal state in agreement with the observation that standard light has no fixed chemical potential and the number fluctuations are infinite. Only in the squeezed state we can associate a gap and a finite ${\cal T}$-matrix which describes the condensate.

In the following it will be convenient to determine the density of condensed particles by the ground state ${\cal T}$-matrix and the gap. The inverse one-particle propagator at zero energy is $G^{-1}(0)=\mu-\Sigma_0=\mu^*$ and in multiple-scattering corrected ${\cal T}$-matrix approximation (\ref{inverse}) we have as well $G^{-1}(0)=n_0 {\cal T}_0$ with the condensate density $n_0$, see (\ref{a4}) in the appendix. We can therefore deduce from the Hugenholtz-Pines theorem (\ref{HP}) the relation \cite{M11_1}
\be
\Delta_0=\Sigma_0-\mu=-n_0 {\cal T}_0
\label{rel}
\ee
where ${\cal T}_0={\cal T}(-\Delta,2\Delta))$ is the ground-state ${\cal T}$-matrix of (\ref{inverse}).
\begin{figure}[]
  \includegraphics[width=8cm]{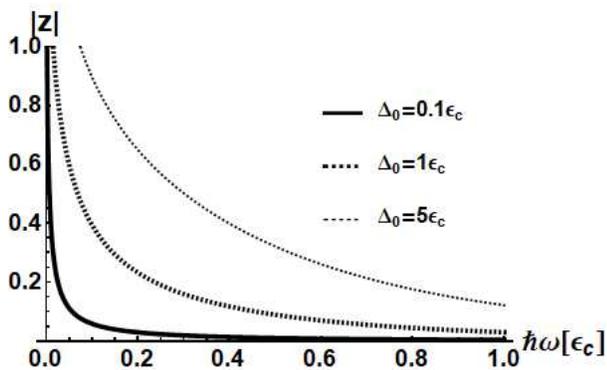}
\caption{The frequency dependence of the squeezing parameter (\ref{z}) with scaling (\ref{ec}).}
\label{z(w)}
\end{figure}
\begin{figure}[]
  \includegraphics[width=8cm]{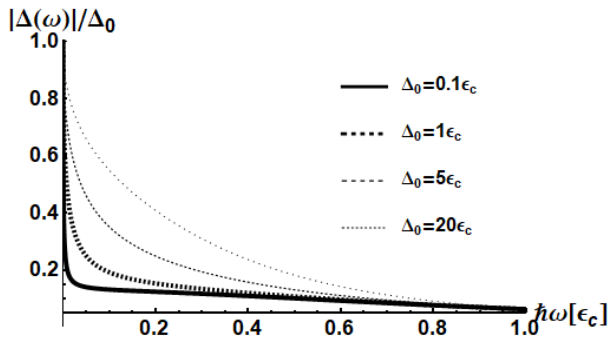}
\caption{The frequency dependence of the equivalent gap (\ref{gap}) with $\Delta_0=\Delta(0)$  with scaling (\ref{ec}).}
\label{d(w)}
\end{figure}

\section{Model of frequency-dependent squeezing}

Now we are going to present a model for the actual gap and squeezing parameter. This will be crucially based on its form of frequency dependence which realization we can suggest here only theoretically.
A scheme to generate frequency-tunable squeezed light has been proposed in \cite{Kum90}. Since its first experimental realization \cite{HK92} one can found more than 170 citations to realize quantum frequency conversion. Other experimental techniques use laser-driven quantum dot regimes to create phonon-induced squeezing \cite{UWRU13}. Such a way high-efficiency squeezed light has been generated between 10 Hz and 100 kHz \cite{Meh18}. An acoustic phonon bath coupled to a quantum dot effectively forms a tunable quantum squeezed reservoir \cite{BGF16}. 
For a recent review of methods see citations therein. Squeezed light can be created even from micromechanical resonators \cite{saf13}. The interaction of wavelength-controlled squeezed photon field with atomic polarization has been investigated in \cite{RIGELS07}. A frequency-tunable source of squeezed light was employed for spectroscopic measurements of atomic cesium \cite{PCK92}.
For a review of application of squeezed light in atomic spectroscopy see \cite{DFS99}. A theoretical prediction to engineer matter interactions using squeezed vacuum can be found in \cite{ZAH17}.

We limit the necessary frequency dependence of the squeezing parameter $z(\omega)$ to render the momentum summation $\sum_q=\int_0^\infty d\omega \omega^2 n_g(\omega)/2\pi^2c^3$ finite. For large frequencies we have for (\ref{np}) with (\ref{uv}) just $n_g(\omega)\sim 2\sinh^2{z(\omega)}$ and we therefore have to demand that $z(\omega)$ falls off faster than
\be
\sinh{z(\omega)}\sim \omega^{-3/2},\qquad\omega\to\infty.
\label{d1}
\ee
For the gap (\ref{gap}) itself we want to demand that $\Delta(0)=\Delta_0\neq 0$ which translates from (\ref{gap})  into
\be
z(\omega)\sim\ln{\Delta_0\over \hbar \omega},\qquad\omega\to 0.
\label{d2}
\ee
For exploratory reasons we choose a model for the frequency-dependence in the form
\be
\sinh{z(\omega)}=\frac{\sqrt{{\Delta_0}+8\hbar\omega}-\sqrt{8\hbar \omega}}{\left(\hbar\omega/\epsilon_c)^2+1\right) \sqrt[4]{8\hbar\omega \left({\Delta_0}+8\hbar\omega\right)}}
\label{ch}
\ee
seen in figure~\ref{z(w)}
which guarantees the expansions (\ref{d1}) and (\ref{d2}). It turns out that the actual form of any chosen model obeying (\ref{d1}) and (\ref{d2}) does not influence the final phase diagram visibly such that it is sufficient to demonstrate it here with the model choice (\ref{ch}).
%\be
%z(\omega)=\left \{\begin{matrix}\frac 1 4 \ln({2 \Delta_0\over \omega})+o(\omega^{1/2})\cr {\Delta_0\over 16 \omega^3}+o(\omega^{-7/2})\end{matrix}\right ..
%\ee
This leads to a frequency dependence of the gap as seen in figure~\ref{d(w)} and the expansion of the gap
\be
\Delta(\omega)=\left \{\begin{matrix}\Delta_0+o(\omega^{1/2})\cr {\Delta_0\epsilon_c\over 8 (\hbar\omega)^2}+o(\omega^{-3})\end{matrix}\right ..
\ee

\begin{figure}[]
  \includegraphics[width=8cm]{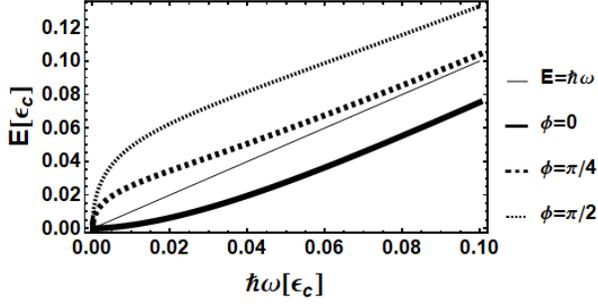}
\caption{The quasiparticle energy (\ref{E1}) for different phases of the squeezing parameter (\ref{z}) with scaling (\ref{ec}).}
\label{e(w)}
\end{figure}

\begin{figure}[]
  \includegraphics[width=8cm]{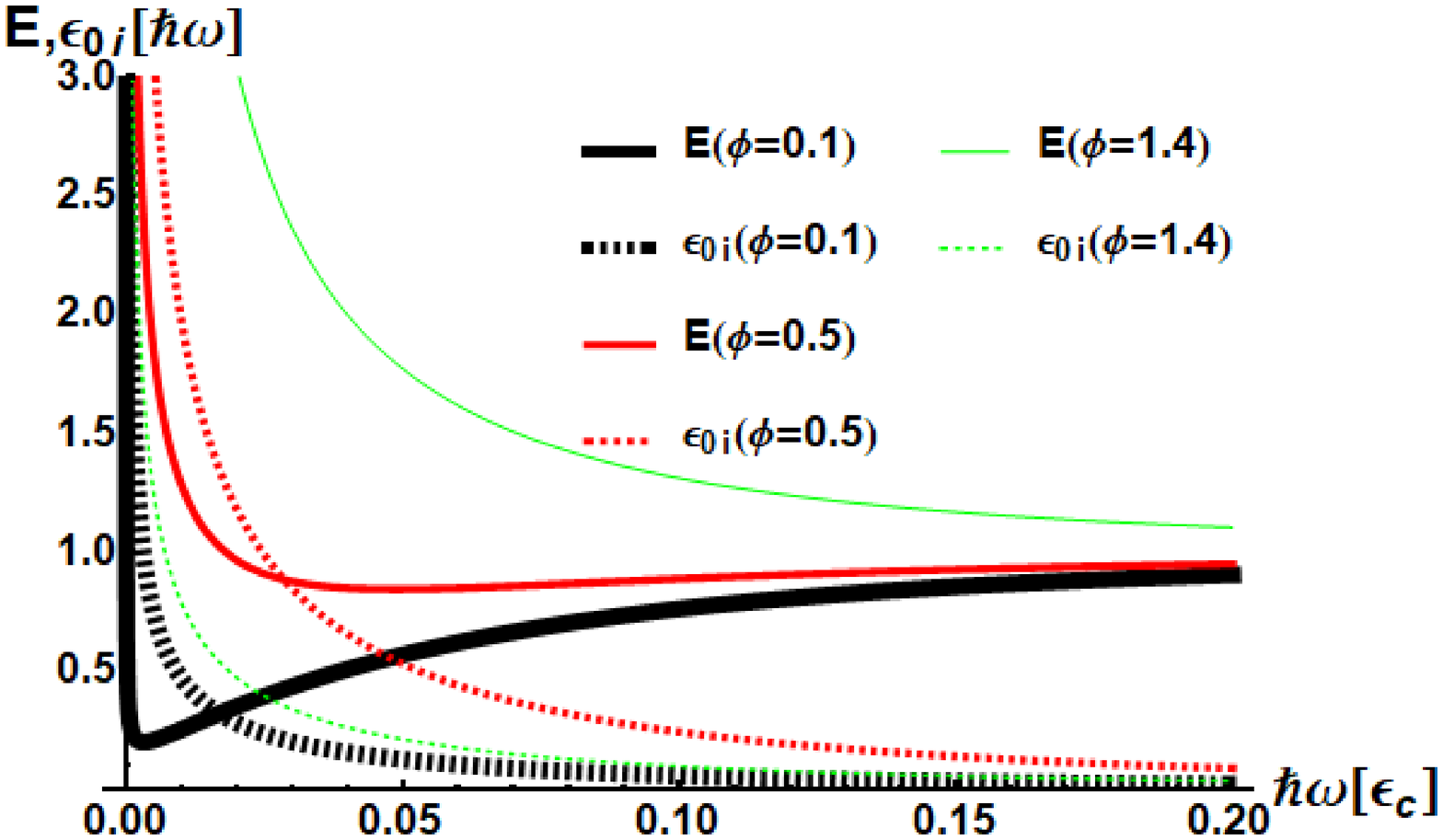}
  \includegraphics[width=8cm]{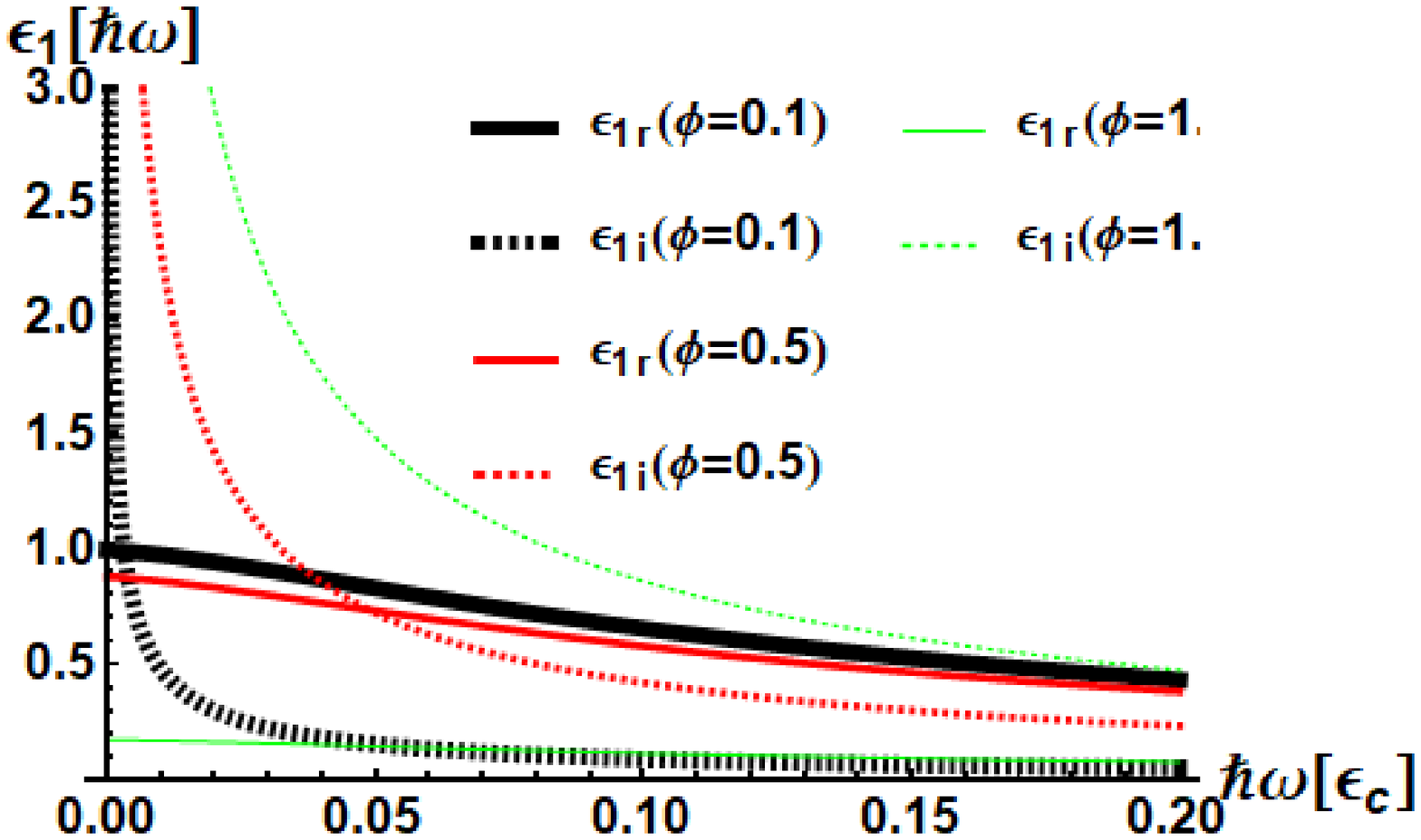}
\caption{Above: The ratio of quasiparticle energy (solid line) according to (\ref{E}) and its inverse lifetime $\epsilon_{0i}=\hbar/\tau$ (dashed) according to (\ref{e0i}) to the frequency versus frequency which is the free quasiparticle energy for various phases. The scale is (\ref{ec}).
Below: The interaction energy and its inverse lifetime according to (\ref{e0i}).
}
\label{Ee0i_w}
\end{figure}

The resulting quasiparticle energy (\ref{E1}) is plotted in figure~\ref{e(w)}. One sees that the actual frequency behaviour is dependent on the complex phase. It can lead to large and smaller values than the free dispersion $E=\hbar\omega$.

We can interpret the complex parts of (\ref{e0i}) as the lifetime of the state, $\epsilon_{0i}=\hbar/\tau$, and the lifetime of the interaction, $\epsilon_{1i}=\hbar/\tau_{\rm int}$. The quasiparticle and interaction energy and their inverse lifetimes are plotted in figure~\ref{Ee0i_w}. The squeezing (interaction) suppresses the free quasiparticle dispersion $\epsilon_{0r}=\hbar \omega$ for low phases $\phi$ but enhances it at larger ones. For frequencies above the crossing point of the quasiparticle energy and inverse lifetime the latter becomes so short that we can consider the system as over-damped. We see that the curves are strongly dependent on the phase $\phi$ of the squeezing parameter which determines the anomalous density (\ref{np}). The corresponding real and imaginary (life time) parts of the interaction energy is seen as well. The low-frequency behaviour is now different from the quasiparticle behaviour and can be understood by the expansions of (\ref{e0i})
\be
E&=&\left \{\begin{matrix}
\sqrt{\hbar\omega \Delta_0\over 2}\sin^2{\phi} + o(\omega^{3/2})
\cr
\hbar\omega -{\cos{2\phi}\over 128 \omega^2(1+(\hbar\omega/\epsilon_c)^2)^2}\Delta_0^2+o(\Delta_0^3)
\end{matrix}\right .
\nonumber\\
\epsilon_{0i}&=&\left \{\begin{matrix}
\sqrt{\hbar\omega \Delta_0\over 8}\sin{2 \phi} + o(\omega)
\cr
{\sin{2\phi}\over 128 \hbar\omega(1+(\hbar\omega/\epsilon_c)^2)^2}\Delta_0^2+o(\Delta_0^3)
\end{matrix}\right .
\nonumber\\
\epsilon_{1r}&=&\left \{\begin{matrix}
\hbar\omega \cos{\phi} + o(\omega^{2})
\cr
{\cos{\phi}\over 8(1+(\hbar\omega/\epsilon_c)^2)}\Delta_0+o(\Delta_0^2)
\end{matrix}\right .
\nonumber\\
\epsilon_{0i}&=&\left \{\begin{matrix}
\sqrt{\hbar\omega \Delta_0\over 2}\sin{\phi} + o(\omega^{3/2})
\cr
{\sin{\phi}\over 8 (1+(\hbar\omega/\epsilon_c)^2)}\Delta_0+o(\Delta_0^2)
\end{matrix}\right .
.
\ee
It shows that the quasiparticle energy starts to deviate from the free dispersion quadratically in the gap and the inverse lifetime appears quadratically with the gap. In contrast, the interaction energy and inverse lifetime starts linearly with the gap. In summary, the squeezing is shown to lead to the same occupations as a complex-valued Hamiltonian which provides lifetime effects. As one can see from (\ref{e0i}) the lifetime is proportional to $1/\sin\phi$ with the phase of the squeezing parameter and can therefore be tuned.

\begin{figure}[h]
  \includegraphics[width=8cm]{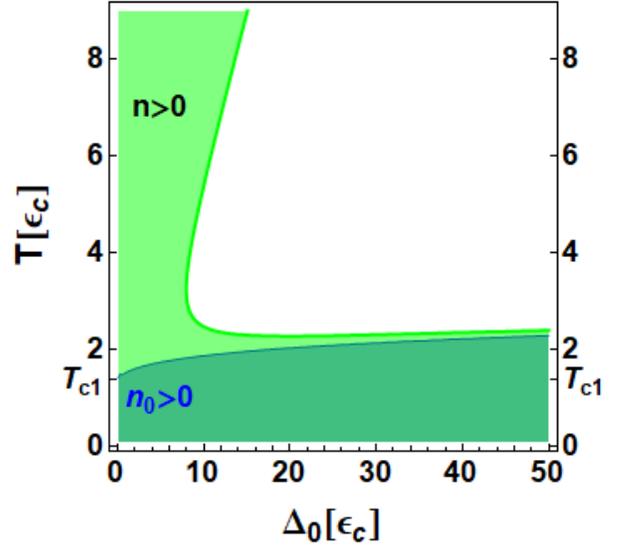}
\caption{The region where the total density and the condensate density are nonzero vs. gap and temperature and scaling (\ref{ec}).}
\label{ngn0_T}
\end{figure}

\section{Results on phase diagram}
Now we calculate explicitly the condensate density from (\ref{rel}) with the help of (\ref{inverse}) and
the total density
\be
n=n_g+n_0=n_g-{\Delta_0\over {\cal T}_0}
\ee
where the correlated density (\ref{fg}) is calculated as 
\be
n_g={1\over 2 \pi^2 c^3} \int\limits_0^\infty d\omega \omega^2 n_g(\omega)
\ee 
with $\mu^*=-\Delta$.
We will use conveniently the parameter $\Delta_0$ to create parametric plots.

\begin{figure}[h]
  \includegraphics[width=8cm]{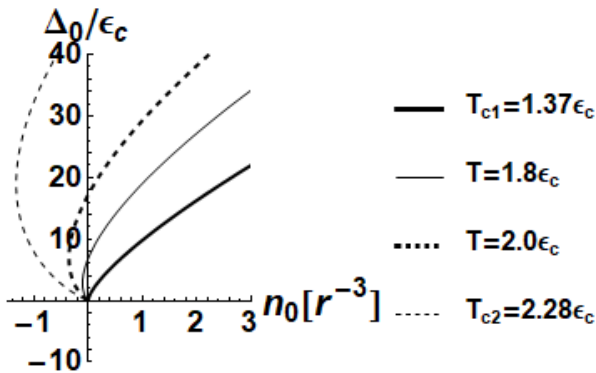}
\caption{The gap versus condensate density for different temperatures in units of (\ref{ec}).}
\label{d_n0}
\end{figure}

\begin{figure}[h]
  \includegraphics[width=8cm]{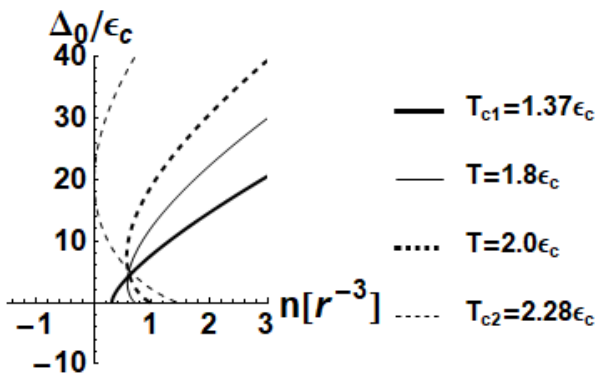}
\caption{The gap versus total density for different temperatures in units of (\ref{ec}).}
\label{d_n}
\end{figure}

The results where the condensate density becomes larger zero is seen in figure~\ref{ngn0_T} which determines the critical density of Bose condensation as function of the gap parameter $\Delta_0$. This condensate density $n_0$ as well as the total density $n$ should be larger zero, of course, which excludes the white area as parameter range. The critical temperature for $n_0=0$ reaches a finite value $T_{c1}$ for vanishing gap. This could suggest that we do have a finite condensate density with vanishing gap. But a closer look at figure~\ref{d_n0} reveals that below $T_{c1}$ the gap starts at zero condensate density and is proportional to the condensate density. In fact for small $\Delta_0$ we have the expansion at low temperatures
\be
n_0r_0^3={\Delta_0\over 2 \pi \epsilon_c}\left (1-{\pi\over 12\epsilon_c^2}T^2+o(T^3)\right )+o(\Delta_0^2).
\ee
Therefore we can state that for temperatures $T<T_{c1}$ we have a positive gap and a condensate density proportional to the gap. From figure~\ref{d_n} we see that in this temperature regime the total density must be larger than the critical one which we call condensate border line. 

\begin{figure}[h]
  \includegraphics[width=8cm]{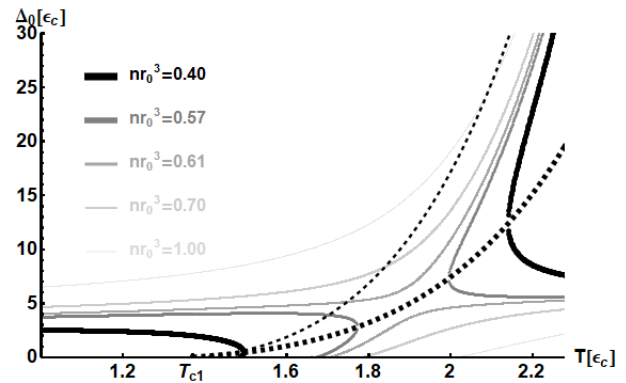}
\caption{The gap parameter versus temperature for various densities. The thick dashed line gives the gap where the minimal density appears, see figure \ref{d_n}. Above the density $n_{c1}$ two gaps appear for one temperature and or densities. The thin dashed line gives the limiting curve above which we have a finite condensate density according to figure~\ref{ngn0_T}.}
\label{double}
\end{figure}

For higher temperatures $T_{c1}<T<T_{c2}$ we see that the gap parameter becomes twofold as a function of the condensate density in figures~\ref{d_n0} and as function of the total density in figure~\ref{d_n}. Considering the total density in figure~\ref{d_n}, the gap starts at a minimal density. This means that for densities above this minimum two gaps are present. Above a critical temperature $T_{c2}$ this minimum would be at negative $n_0$ and we have a missing range in the gap.

The temperature dependence of the gap is plotted in figure~\ref{double} for various densities. The thick broken line gives the value of the gap versus temperature where the minimal density appears according to figure~\ref{d_n}. The thin line gives the range above which we only have a positive condensate density $n_0$ according to figure~\ref{d_n0}. This shows that in the range $T_{c1}<T<T_{c2}$ we have two gaps, however only the larger one leads to a positive condensate density.

This behaviour is summarized in the phase diagram of figure~\ref{phases}. 
The shaded area indicates the range where we have a finite condensate density. for temperatures below $T_{c1}$ this agrees with the range of a finite gap. For the temperature range between $T_{c1}$ and $T_{c2}$ the minimum of figure~\ref{d_n} leads to an area of two gaps above the limiting curve. In this region we do have only a finite condensate density above the borderline. Above $T_{c2}$ finally the two gaps appear already at zero density but the finite condensate density we have again above the borderline.

Additionally we plot the general critical curve of figure~\ref{Tc} in chapter~\ref{pha} as thick line in figure~\ref{phases}. It starts exactly at $(T_{c1},n_{c1})$ but shows a bending over and vanishes at $n_{c2}$ corresponding to $T_{c2}$. The latter observation is nontrivial. We have obtained this curve as inverse series expansion of the ${\cal T}$-matrix in chapter~\ref{pha} which yields higher-order approximation of correlations than the ${\cal T}$-matrix itself \cite{FKYSOI95}. On the other hand we had assumed there just any interaction and have shown that this line appears as universal phase border. Concluding we can consider the possible range of Bose condensation of squeezed light as the area limited by the densities $n_{c1}$ and $n_{c2}$ as well as temperatures below the limiting curve.

\begin{figure}[h]
  \includegraphics[width=8cm]{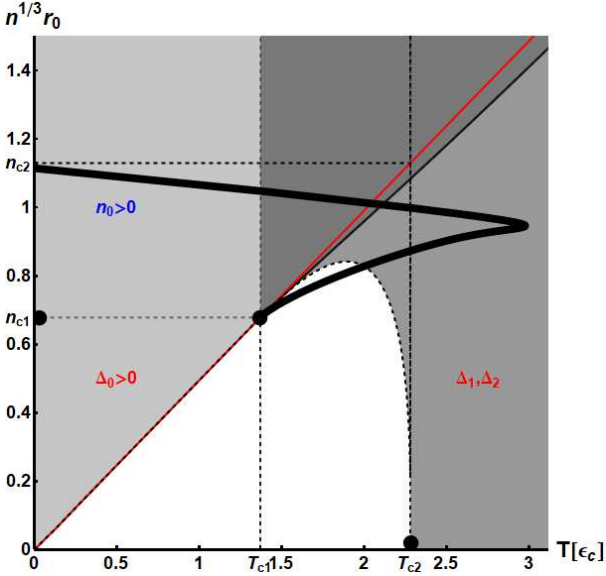}
\caption{The phase diagram in the density versus temperature plot in units of (\ref{ec}). The region where the gap is nonzero (shaded) includes the region where the condensate density is nonzero (dark gray) which lies above the straight borderline. The curve between the two right dots are the minima of the density with respect to the gap of figure~\ref{d_n}. In this range the straight line slightly above the borderline  would be the line of zero gap in figure~\ref{d_n}. For temperatures higher than  $T_{c1}$ two gaps appear. The critical curve from figure~\ref{Tc} is given by the thick line for comparison. It starts exactly at $\{T_{c1},n_{c1}\}$  and is observed to end at $\{0,n_{c2}\}$. }
\label{phases}
\end{figure}

\section{Summary}

We have investigate the general phase diagram for the Bose-Einstein condensation of interacting light with an assumed effective chemical potential. By employing the divergence of the in-medium scattering range, the general critical line is found by the inversion expansion method. The curves show a back-bending in that two different critical densities appear at one critical temperature with an upper critical density and a maximal critical temperature. The assumed chemical potential can be generated e.g. by the interactions with the surrounding Dye molecules.  As new proposal here such interacting light might be realized by a proper squeezing. To this end an equivalence of squeezing with a complex Bogoliubov transformation is found with the help of which an effective gap is derived. The squeezed light is found to be represented by an interacting and gapped Bose gas leading to a finite chemical potential and finite lifetime of the excitation. The resulting phase diagram in the density and temperature shows regions with one and with two gaps and a region of finite condensate density. It is found to be consistent with the general phase diagram above. The Bogoliubov transformation establishes an equivalence between squeezing and a complex-valued Hamiltonian with respect to the occupation numbers. For a non-zero phase of squeezing this leads to finite life-times of the states. Concluding we interpret the proposed possibility of Bose-Einstein condensation of light by squeezing as being accompanied by finite lifetimes which can be tuned by the phase of the squeezing parameter such that the effect might be stable enough to be observed. So far the present investigation is a purely theoretical suggestion and probably far from experimental realization.

\appendix

\section{Multiple-scattering corrected T-matrix in the condensate phase}

The Bethe-Salpeter equation for the retarded two-particle T-matrix (\ref{T}) can be solved for separable interaction $V_{12}=\lambda g_1g_2$
\be
T^R_{12}&=&V_{12}+\sum\limits_{34} V_{13}{\cal G}^R_{34}T^R_{32}
\nonumber\\
&=&{\lambda g_1 g_2\over 1-\lambda \sum\limits_{34} g_3{\cal G}^R_{34}g_4}.
\nonumber\\
T^R_{K,p}&=&{\lambda g_{p}^2 \over 1-\lambda \int {d^3q\over (2\pi \hbar)^3} g_q^2{\cal G}^R_{K,q}}
\label{a1}
\ee
where we give the dependence on center-of mass momentum $K$ and difference momentum $p$ in the last line.
The two-particle propagator in time
\be
{\cal G}^R_{12}(t,t')=-i\Theta(t-t') \left [{\cal G}^>_{12}(t-t')-{\cal G}^R_{12}(t-t')\right ]
\ee
is Fourier transformed into frequency and we take according to chapter~\ref{tcorr} one quasiparticle propagator, 
\be
G_1^\gtrless=\begin{pmatrix}1+f_{\epsilon_1}\cr f_{\epsilon_1}\end{pmatrix}2\pi \delta (\hbar\omega-\epsilon_1),
\label{a2}
\ee
and one subtracted propagator, see (\ref{polebose}),
\be
G_\2^\gtrless&=& \begin{pmatrix}1+f_{E_2}\cr f_{E_2}\end{pmatrix}\left (1+{\epsilon_2\over E_2}\right )\pi \delta (\hbar\omega_2-E_2)
\nonumber\\
&&
-\begin{pmatrix}f_{E_2}\cr 1+f_{E_2}\end{pmatrix}\left (1-{\epsilon_2\over E_2}\right )\pi \delta (\hbar\omega_2+E_2)
\ee
into account with the free energy (\ref{eps}) and quasiparticle energy (\ref{E1}). The result reads
\ba
{\cal G}^R_{12}(\omega)={1\!+\!f_{\epsilon_1}\!+\!f_{E_2}\over \hbar \omega\!-\!E_2\!-\!\epsilon_1}\!\left (\!\frac 1 2\!+\!{\epsilon_2\over 2E_2}\!\right )
\!+\!
{f_{\epsilon_1}\!-\!f_{E_2}\over \hbar \omega\!+\!E_2\!-\!\epsilon_1}\!\left (\!\frac 1 2\!-\!{\epsilon_2\over 2 E_2}\!\right ).
\end{align}
For zero center-of-mass momenta $K=0$ one has $\epsilon_1=\epsilon(\frac K 2+p)=\epsilon_2=\epsilon(\frac K 2-p)$ and for $\omega=0$ one obtains the usual pairing expression
\be
{\cal G}^R_{12}(0)=-{1+2 f_{E_2}\over E_2}.
\ee
This leads with the replacement of $\lambda=\lambda(r_0)$ according to the discussion after (\ref{r}) to the ${cal T}$-matrix in the condensate phase
\be
{\cal T}(-\Delta_0,2\Delta_0)={2\pi \epsilon_c r_0^3\over \int\limits_0^\infty {d\Omega\over 2 \pi} \Omega^2\left ({\coth{E\over 2 T}\over E/\epsilon_c}-\frac 1 \Omega\right )-1}
\ee
where $\Omega=c q/\epsilon_c$.

The zero of the denominator (\ref{a1}) provides the determination of the gap phase. Since we expect a condensate density at zero momentum, we have to split this density off the occupation (\ref{a2})
\be
G_1^\gtrless=G_{12\rm normal }^\gtrless+n_0 (2\pi\hbar)^3 \delta (p_1)2\pi \delta (\hbar\omega-\epsilon_1).
\label{a3}
\ee
In the condensate regime, $\epsilon_1=-\mu^*$, this provides in the two-particle propagator (\ref{a3}) the explicit term
\be
{\cal G}^R_{12}(\omega)={\cal G}^R_{12\rm normal }(\omega)+n_0 (2\pi\hbar)^3 \delta (p) {1\over \mu^*}.
\ee
Therefore the pole condition of (\ref{a1}) reads
\be
1-\lambda {n_0 g_0^2\over\mu^*}-\lambda \int\limits_{q\ne 0} {d^3q\over (2\pi \hbar)^3} g_q^2{\cal G}^R_{K,q}=0
\ee
or rewritten
\be
\mu^*={\lambda g_0^2 n_0\over 1-\lambda \int\limits_{q\ne 0} {d^3q\over (2\pi \hbar)^3} g_q^2{\cal G}^R_{K,q}}={\cal T}_0 n_0.
\label{a4}
\ee
This relation together with the Hugenholtz-Pines relation (\ref{HP}) will serve to determine $n_0(\Delta)$.

\acknowledgements
The author wants to thank the anonymous referees for many elaborate and helpful comments.

\bibliography{entropy,bose,kmsr,kmsr1,kmsr2,kmsr3,kmsr4,kmsr5,kmsr6,kmsr7,delay2,spin,spin1,refer,delay3,gdr,chaos,sem3,sem1,sem2,short,cauchy,genn,paradox,deform,shuttling,blase,spinhall,spincurrent,tdgl,pattern,zitter,graphene,quench,msc_nodouble,iso,march,weyl,polariton,squeeze}
\bibliographystyle{apsrev}

\end{document}